\documentclass{aa}  

\usepackage{graphicx}
\usepackage{color}

\usepackage[varg]{txfonts}

\begin{document}

\title{Consequences of a strong phase transition in the dense
matter equation of state for the rotational evolution of neutron stars}
\authorrunning{Bejger et al.}
\titlerunning{Rotating high-density twin neutron stars}

\author{M. Bejger$^a$, D. Blaschke$^{b,c,d}$, P. Haensel$^a$, J. L. Zdunik$^a$, M. Fortin$^a$}
\institute{
$^a$Nicolaus Copernicus Astronomical Center, Polish Academy of Sciences, Bartycka 18, PL-00-716 
Warsaw, Poland
\\ \email{bejger@camk.edu.pl, haensel@camk.edu.pl, jlz@camk.edu.pl, fortin@camk.edu.pl}
\\[3mm]
$^b$Institute of Theoretical Physics, University of Wroclaw, pl. Maxa Borna 9, 
PL-50-204 Wroclaw, Poland
\\
$^c$Bogoliubov Laboratory for Theoretical Physics, JINR Dubna, ul. Joliot-Curie 6, 
RU-141980 Dubna, Russia 
\\
$^d$National Research Nuclear University (MEPhI), Kashirskoe Shosse 31, 
RU-115409 Moscow, Russia 
\\ \email{blaschke@ift.uni.wroc.pl}
}

\date{Received ...; accepted ...}

\abstract
% context
{}
% aims
{We explore the implications of a strong first-order phase transition region in
the dense matter equation of state in the interiors of rotating neutron stars,
and the resulting creation of two disjoint  families of neutron-star
configurations (the so-called high-mass twins).}
% methods
{We numerically obtained rotating, axisymmetric, and stationary stellar configurations in the framework of general relativity, and studied their global parameters
and stability.}
% results
{The instability induced by the equation of state divides stable neutron star
configurations into two disjoint families: neutron stars (second family) and
hybrid stars (third family), with an overlapping region in mass, the high-mass
twin-star region.  These two regions are divided by an instability strip.
Its existence has interesting astrophysical consequences for rotating neutron
stars.  We note that it provides a natural explanation for the rotational
frequency cutoff in the observed distribution of neutron star spins, and for
the apparent lack of back-bending in pulsar timing. It also straightforwardly
enables a substantial energy release in a mini-collapse to another neutron-star
configuration (core quake), or to a black hole.}
% conclusions
{}
\keywords{stars: neutron -- pulsars -- equation of state}
\maketitle
%///////////////////////////////////////////////////////////////////////
\section{Introduction}
%///////////////////////////////////////////////////////////////////////

Recent observations of the high-mass pulsars PSR J1614-2230
\citep{Demorest2010,Fonseca2016} and PSR J0348+0432 \citep{Antoniadis2013}
with masses $M\, {\approx}2 M_\odot$ has motivated the nuclear and particle
physics communities to deepen their understanding of the equation of state
(EOS) of high-density matter and of the possible role of exotic states of
matter in neutron star (NS) interiors. 
For further constraints on the high-density EOS from NS and heavy-ion collision 
experiments see \cite{Klahn:2006ir} and \cite{Klahn:2011au}.

Measurements of high masses of NSs do not immediately imply that very high
densities prevail in their cores, meaning that a transition to exotic forms of
matter (hypernuclear matter, quark matter) has to be invoked.  Of
two models for the high-density EOS with the same stellar mass, the stiffer
model will lead to a lower central density but to a larger radius. Therefore,
radius measurements for high-mass pulsars are of the utmost importance. 

Current radius measurements are controversial. 
Determinations of the radius $R$ range from about 9 km \citep{Guillot2013} to 15 km \citep{Bogdanov2013},
but one must be aware of possible systematic flaws (see, e.g., \citealt{Heinke2014,EH16}); 
for a recent critical assessment see, for example, \cite{Fortin2015,Miller2016,Haensel2016}.
At a gravitational mass of ${\sim}2 M_\odot,$ the range of radius values mentioned
above would correspond to a range of central densities of the compact star
between $2.5 - 6~n_0$, where $n_0=0.15$ fm$^{-3}$ is the nuclear
saturation density.

Several observational programs for simultaneous measuring of
pulsar masses and radii are currently in preparation: the Neutron star Interior Composition
ExploreR (NICER, \citealt{NICER}), the Square Kilometer Array (SKA,
\citealt{SKA}), Athena \citep{Athena+}, and possibly, a LOFT-size mission
\citep{LOFT}. Thus there is hope that in the near future it will be possible to
reconstruct  the cold NS matter EOS
$P(\varepsilon)$ (here $P$ is pressure and $\varepsilon$ is energy density) within the measurement errors
from the measured $M(R)$ relation by means of inverting the
Tolman--Oppenheimer--Volkoff (TOV, \citealt{Tolman1939,OppenheimerV1939})
equations. In addition, specific proposals to use gravitational waves
(GWs)  for measuring the NS radius from either the observations of the inspiral
(e.g., \citealt{Bejger2005,DamourNV2012}) or post-merger waveforms (e.g.,
\citealt{BausweinSJ2015}) were presented.

The systematic investigation of a wide class of hybrid stars with varying
stiffness of hadronic matter at high densities and the possibility of quark
matter with varying high-density stiffness has revealed interesting findings
\citep{Alvarez2016}. Possible future measurements of radii of recently
discovered high-mass stars \citep{Demorest2010,Antoniadis2013} would select a
hybrid EOS with a strong first-order phase transition if the outcome of their
radius measurement were to show a difference of about 2 km with significance.
Such a possibility has been suggested earlier on the basis of a new class of
hybrid star EOS that fulfill the generic condition that the baryonic EOS is
strongly stiffened at high densities, for instance, by effects of the Pauli exclusion
principle (quark exchange interaction between baryons), and a strong first-order
deconfinement phase transition requiring sufficiently soft quark matter at the
transition between baryonic and quark phases, ${\rm B\longrightarrow Q}$. The
quark matter EOS has, however, to stiffen quickly with increasing density so that
immediate gravitational collapse that is due to the transition does not occur, allowing
stable hybrid stars to exist.

Such a solution for hybrid stars, which form a third family of compact stars
that are disconnected from the baryonic branch of compact stars, is very
interesting for the possible observational verification of specific
features of phase transition to quark matter in NS cores. A sharp
first-order phase transition between pure B and Q phases, occurring at constant
pressure $P_{_{\rm BQ}}$, is associated with an energy density jump from
$\varepsilon_{_{\rm B}}$ to $\varepsilon_{_{\rm Q}}$ ($\varepsilon$ is the  
energy density including the rest energy of particles). Then, a general necessary 
condition for the existence of a disconnected family of NSs with Q-phase cores is
$\varepsilon_{_{\rm Q}}/\varepsilon_{_{\rm B}}>\lambda_{\rm crit}$, with
$\lambda_{\rm crit}=\frac{3}{2}(1+P_{_{\rm BQ}}/\varepsilon_{_{\rm B}})$
\citep{Seidov1971}. The second term in the brackets comes from general 
relativity. The condition implies that hybrid stars with small quark cores are
unstable to radial perturbations and collapse into black holes (BH), and
therefore $\varepsilon_{_{\rm Q}}/\varepsilon_{_{\rm B}}>\lambda_{\rm
crit}>\frac{3}{2}$ implies the existence of a separate hybrid stars branch (a detailed review of the structure and stability of NSs with phase transitions in their cores is given in Sect.~7 of \citealt{HaenselPY2007}). This corresponds
to a separate and small $R$ segment of the $M(R)$ relation. High mass and small
radii imply high spacetime curvature and strong gravitational pull, resulting
in a relatively flat (nearly horizontal) $M(R)$ segment. This hybrid star
branch is characterized by a narrow range of $M$, a broad range of $R$, with
$M$ weakly increasing with decreasing $R$, up to a very flat $M$ maximum.
These features of the hybrid-star branch result in specific  observational
signatures  of a strong B$\longrightarrow$Q first-order phase transition 
in NS cores. Moreover, these generic properties of the hybrid-star branch
indicate a relative ''softness'' of their configurations with respect to
perturbations that are due to rotation and oscillations, for
example. All these generic
features are studied in the present paper, using an illustrative example of
an advanced  EOS composed of a stiff baryonic segment and a strong first-order
phase transition to the quark phase (Sect.~3). 
 
The ${\rm B\longrightarrow Q}$ phase transition might occur smoothed
within a finite pressure interval through a mixed BQ phase layer. However, the
interplay of the surface tension at the B-Q interface and charge screening  of
Coulomb interaction (in a mixed state, B and Q phases are electrically charged)
make the mixed-state layer very thin \citep{Endo2006} so that the key features
of the hybrid-star branch remain intact \citep{Alvarez-Castillo2015:2014dva}. 
 
Stable branches of static configurations in the $M-R$ plane have very 
specific generic features (see, e.g., \citealt{Benic2014}). 
The high-mass baryon branch is very steep, not only nearly vertical, but even with $R$ 
increasing with $M$, a feature characteristic of very stiff NS cores. 
The hybrid stable twin-branch is predicted to be flat, nearly horizontal with very 
broad maximum. Measuring radii of ${\sim}2 M_\odot$ stars, which span a wide range of values 
from about $12$ to $15$ km, will clearly indicate a hybrid branch, and if 
moreover the radii for standard NS masses, $1.2 - 1.6 M_\odot$,  are roughly
constant at approximately $14$ km,  then the evidence for two  distinct families  that are separated as a result of a strong 
first-order phase transition  would be quite convincing. The range of central 
pressures for NSs of different radii at $2 M_\odot$, which may bear a connection 
to the universal hadronization pressure found in heavy-ion collisions, is discussed 
in \citet{Alvarez2016b}. 

This possibility of finding observational evidence for a first-order phase
transition in NS cores from the phenomenology of $M-R$ characteristics of NS
populations offers the chance to answer to the currently controversial question 
of whether a critical endpoint of first-order phase transitions in the QCD phase diagram exists 
\citep{Alvarez-Castillo:2013cxa,Blaschke:2013ana}.

Since the nature of the QCD transition at vanishing baryon density is beyond
doubt identified as a crossover in two independent lattice QCD simulations at
the physical point \citep{Bazavov:2014pvz,Borsanyi:2013bia}, the evidence for a
first-order phase transition at zero temperature and finite baryon density
necessarily implies the existence of a critical endpoint (CEP) of first-order
phase transitions in the QCD phase diagram (we note that
there are theoretical conjectures about a continuity between hadronic and quark
matter phases at low temperatures and finite densities in the QCD phase diagram
\citep{Schafer:1998ef,Hatsuda:2006ps,Abuki:2010jq}; 
if, however, there were observational evidence for a horizontal branch in the $M-R$ diagram of
compact stars, these considerations would become obsolete). The very
existence of such a CEP is a landmark for identifying the universality class of
QCD, and due to its importance for model building and phenomenology, a major
target of experimental research programs with ultra-relativistic heavy-ion
collisions at BNL RHIC (STAR beam energy scan, \citealt{Stephans:2006tg}) and
CERN SPS (NA49 SHINE, \citealt{Gazdzicki:2006}), in future at NICA in Dubna
\citep{Sissakian:2006dn} and at FAIR in Darmstadt \citep{FAIR}.    

First examples for microscopically founded hybrid star EOS that simultaneously
fulfill the above constraints for the existence of a disconnected hybrid star
branch and for a high gravitational mass of about  $2~M_\odot$ (which implies
the existence of so-called high-mass twin stars) have been given in
\citet{Blaschke:2013ana}.  The recent systematic investigation of high-mass
twin stars in \citet{Alvarez-Castillo2015:2014dva}  is based on the EOS
developed in \citet{Benic2014}, which joins a relativistic density functional
for nuclear matter with nucleonic excluded volume stiffening with a
Nambu--Jona-Lasinio (NJL) type model for quark matter that provides a
high-density stiffening as a result of higher order quark interactions. These examples
belong to a new class of EOS that can be considered as a realization of the
recently introduced three-window picture for dense QCD matter 
\citep{Kojo:2015fua} as a microscopic foundation for the hybrid EOS that was
conjectured by \citet{Masuda:2012ed}, suggesting a crossover construction
between hadronic and quark matter. These three windows depicted in Fig.~1 of
\citet{Kojo:2015fua} are characterized by the following:

\begin{itemize}
\item \underline{at densities $n_{\rm b} < 2~n_0$:} occasional quark exchange between
separated and still well-defined nucleons, leading to quark Pauli blocking
effects in dense hadronic matter \citep{Ropke:1986qs} that can be modeled by a
hadronic excluded volume or by repulsive short-range interactions,  
\item \underline{densities $2~n_0<n_{\rm b}< 5~n_0$:} multiple quark exchanges that
lead to a partial delocalization of the hadron wavefunctions and to the
formation of multi-quark clusters and a softening of the EOS by an attractive
mean field,
\item \underline{densities $n_{\rm b}>5~n_0$:} baryon wave functions overlap and
quarks become delocalized; a Fermi sea for strongly interacting quark matter forms
with higher order repulsive short-range interactions, for example, by multipomeron
exchange \citep{Yamamoto:2015lwa} or nonlinear quark interactions
\citep{Benic:2014iaa}.
\end{itemize}

A still open and controversial question in this context is whether chiral
symmetry restoration and deconfinement transition (which both coincide on the
temperature axis according to lattice QCD simulations) would also occur
simultaneously in the dense matter at zero temperature. If this were not
the case, then there would be room for a more complex structure of the QCD phase
diagram, for instance, with a triple point that is either due to a quarkyonic matter phase
(light quarks confined in baryons that form parity doublets,
\citealt{McLerran:2007qj}) or a massive quark matter phase
\citep{Schulz:1987qg} for which there is circumstantial evidence from particle
production in ultrarelativistic heavy-ion collision experiments
\citep{Andronic:2009gj}. More circumstantial evidence for a region of strong
first-order phase transitions in the QCD phase diagram comes from the baryon
stopping signal in the energy dependence of the curvature of the net proton
rapidity distribution at midrapidity for energies in the intermediate range
between former AGS experiments and the NA49 experiment at the CERN SPS
\citep{Ivanov:2012bh}. This signal has been proven to be quite robust under
different experimental constraints \citep{Ivanov:2015vna} and against hadronic 
final state interactions \citep{Batyuk2016}. 

On the basis of this discussion, the new class of EOS with a three-window
structure provides the theoretical background for a {\it \textup{strong first-order
phase transition}}, the phenomena in the energy scan of heavy-ion collision
experiments, and  the creation of two disjoint families of NS configurations.
Astrophysical observations of the {\it \textup{NS twins}} with drastically different
core compositions may be regarded as a manifestation of these dense matter EOS
features   in different regions of the QCD phase diagram and under different
physical conditions. 

In the present work we add to the discussion of high-mass twin
stars a detailed investigation of their properties under rigid rotation. This
is because  we expect a strong response of both branches to rotation. First, the
high stiffness of the high-density EOS of the baryonic branch  results in large
radii: $R\;{\sim}15$ km at $M\;{\sim}2\,M_\odot$, so that the effect of the
centrifugal force will be large. Second, the stable static hybrid branch is flat, which 
makes it particularly susceptible to the effects of rotation: 
the margin of stability along this branch is narrow. 

The neutron star instability induced by a strong phase transition in the EOS was
studied in detail by \citet{Zdunik2006}, who conjectured that the character of
stability is not changed by the rotation rate of the star (disjoint families
remain separated at any rotation rate). This question will be of particular
interest for discussing compact star phenomena tied to the evolution of their
rotational state, which eventually  ends by collapse into a black hole (e.g.,  
\citealt{FalckeR2014}). 

The general idea of the present work is to illustrate generic features of a
class of high-density quark phase transition EOSs using an exemplary EOS, in
order to discuss the regions of stable and unstable configurations related to a
strong first-order phase transition. For illustration, we use one of the
EOS that was recently developed by \citet{Benic2014}.

The article is structured as follows. In Sect.~\ref{sect:methods} we describe the methods we used
to calculate the rotating configurations, stressing
particularly the need for high precision of numerical simulations. Precision is
particularly important for testing stability criteria of stationary rotating
configurations, which are formulated there. Section \ref{sect:results} starts
with a brief presentation of the EOS that we use to illustrate generic
properties of the rotating high-mass twins. Then we construct families of
rotating configurations, assess their stability, and classify the regions of
instability and their generic features. Discussion of the results in
Sect.~\ref{sect:discussion} involves evolutionary considerations, potential
scenarios leading to observational manifestations of massive-twin case,
including dynamical  phenomena triggered by the instabilities and their
possible astrophysical appearances. The final part of
Sect.~\ref{sect:discussion} presents the conclusions. 

%///////////////////////////////////////////////////////////////////////
\section{Methods}
\label{sect:methods}
%///////////////////////////////////////////////////////////////////////

In order to analyze the astrophysical consequences of the above-mentioned
specific type of EOS, we have obtained rigidly rotating, stationary, and
axisymmetric NS configurations by means of the numerical library {\tt
LORENE}\footnote{\tt http://www.lorene.obspm.fr} {\tt nrotstar} code, using the
3+1 formulation of general relativity of \citet{BonazzolaGSM1993}, and 
employing the multidomain pseudo-spectral decomposition (three domains inside the star). 
The accuracy is controlled by a 2D general-relativistic virial theorem
\citep{BonazzolaG1994} and for the results presented here is typically on the
order of $10^{-7}$. The high accuracy provided by the spectral method
implementation of {\tt LORENE} and a general low numerical viscosity of
spectral methods is particularly suitable for studying the stability of NS
models. 

The evanescent error behavior of the solutions can be employed
by
expanding the number of coefficients to obtain the relevant values practically
up to machine precision. In this context, we recall that a global
parameter that is strictly conserved during the evolution of an isolated 
NS is its total baryon number (baryon charge) $A_{\rm b}$. 
Instead of $A_{\rm b}$ it is convenient to use in relativistic astrophysics 
the\textup{ baryon mass} (or rest mass\textup{{\it })} of the NS defined by $M_{\rm b}=A_{\rm
b}m_{\rm b}$, where $m_{\rm b}$ is a suitably defined baryon mass. $M_{\rm b}$
is the total mass of $A_{\rm b}$  non-interacting baryons; it is easy to extend 
these definitions to NS cores built of quarks (three quarks contribute +1 to the
baryon number of NS). In our calculations we follow the LORENE unit convention
and use a value of the mean baryon mass $m_{\rm b}=1.66\times 10^{-24}$ g.  Other
definitions of $m_{\rm b}$, suitable for specific applications in NS and
supernova physics, are discussed in Sect.~6.2 of \citet{HaenselPY2007}. We note
that while $M_{\rm b}$ is strictly constant in spinning down or cooling
isolated NS, their gravitational mass is changing.  We are in general
interested in obtaining accurate values of the gravitational mass $M$, the
baryon mass $M_{\rm b}$ and, the total angular momentum $J$ that are the
properly defined functionals of the stellar structure and EOS suitable to study
instabilities (for more details and a discussion comparing the 3+1
formulation with the slow-rotation formulation see the recent review of the
stability of rotating NSs with exotic cores, \citealt{Haensel2016}).  

In the following we use the method described in
\citet{Zdunik2004,Zdunik2006}, who studied, among other things, the {\it
\textup{back-bending}} phenomenon proposed for NSs in \citet{Glendenning1997}.
Back-bending, a temporary spin-up of an isolated NS that is decreasing its
total angular momentum $J$ by the dipole radiation, for example, can be robustly
quantified by analyzing the extrema of the baryon mass $M_{\rm b}$ along the
lines of constant spin frequency $f$, that is, the rate of the baryon mass $M_{\rm
b}$ change with respect to a central EOS parameter $\lambda_c$ (central
pressure $P_c$, e.g.,) changing along $f=const.$ lines. The condition for
back-bending to occur is 
\begin{equation}
\left(\frac{\partial M_{\rm b}}{\partial\lambda_c}\right)\bigg\rvert_f < 0 
\label{eq:back-bend}
\end{equation}
(see also \citealt{Haensel2016}, Sect.~7 for detailed description).
Nevertheless, the susceptibility to back-bending {\it \textup{does not mean}} that the
NS is indeed unstable. In order to study the regions of true instability, we
use the turning-point theorem formulated for rotating stellar models by
\cite{Sorkin1981,Sorkin1982,FriedmanIS1988}, which states that the sufficient
condition for the change in stability corresponds to an extremum of the
gravitational mass $M$, or the baryon mass $M_{\rm b}$ at fixed $J$: 
\begin{equation}
\left(\frac{\partial M}{\partial\lambda_c}\right)\bigg\rvert_{J}=0,\qquad 
\left(\frac{\partial M_{\rm b}}{\partial \lambda_c}\right)\bigg\rvert_{J}=0,
\label{eq:inst_cond1}
\end{equation}
or equivalently, to an extremum of $J$ at fixed either $M$, or $M_{\rm b}$: 
\begin{equation}
\left(\frac{\partial J}{\partial \lambda_c}\right)\bigg\rvert_{M} =0,\qquad 
\left(\frac{\partial J}{\partial \lambda_c}\right)\bigg\rvert_{M_{\rm b}} =0. 
\label{eq:inst_cond2}
\end{equation}  
In the following, we illustrate the generic features of the third family
of compact stars with one of the EOSs developed by \citet{Benic2014}. The baryon
phase EOS is based on the relativistic mean-field (RMF) model DD2 of \citet{Typel2010DD2}. 
It fits the semi-empirical parameters of nuclear matter at saturation
well, and features 
density-dependent coupling constants. In order to obtain a
high-density stiffening of B phase, which is necessary to yield a strong
B$\longrightarrow$Q phase transition, while fulfilling $M_{\rm max}>2\;M_\odot$
for the hybrid Q branch, the  DD2 model is modified using  excluded volume (EV)
effects that are included to account for the finite size of baryons, resulting in a
DD2+EV model for the baryon phase. The EV effect was included without altering
the good fit of DD2 to semi-empirical nuclear matter parameter values.  The quark
phase description is based on the Nambu--Jona-Lasinio density-dependent model,
with dimensionless coupling constants $\eta_2=0.12$ and $\eta_4=5$. The
complete EOS with a strong  first-order phase transition ${\rm B\longrightarrow
Q}$ was obtained using Maxwell construction.  We analyze the features of
sequences of configurations along the lines of fixed baryon mass $M_{\rm b}$
and total angular momentum $J$.  

%///////////////////////////////////////////////////////////////////////
\section{Results}
\label{sect:results}
%///////////////////////////////////////////////////////////////////////

The static configurations of NSs for the sample EOS have been studied in 
\citet{Benic2014}. The basic stellar parameters were the gravitational mass $M$ and
the circumferential radius $R$. To analyze the stability of rotating stars, it is
more convenient to consider  the baryon mass $M_{\rm b}$ (also called the rest
mass) of the star and the equatorial circular radius $R_{\rm eq}$ (see, e.g.,
\citealt{Haensel2016}). Generally, the study of rapidly rotating star
configurations offers advantages for the investigation of phase transitions in
NS interiors because as a function of the angular momentum (rotation frequency),
which is an additional parameter, the profile of the density distribution and
therefore also the interior composition would change, which is
expected to allow 
for observational signatures in the course of the rotational evolution of the
star (for early examples, see \citealt{Glendenning1997,Chubarian:1999yn}). The
spin frequency range we studied spans the astrophysically relevant range from
$f=0$ Hz (static configurations, corresponding to the solutions of the
Tolman-Oppenheimer-Volkoff equations), to $f=1000$ Hz (i.e., much higher than the
frequency of $716$ Hz of the most rapid pulsar known to date, PSR J1748-2446ad of
\citealt{Hessels2006}). 

%-----------------------------------------------------------------------
\begin{figure}[ht]
\resizebox{\columnwidth}{!}
{\includegraphics{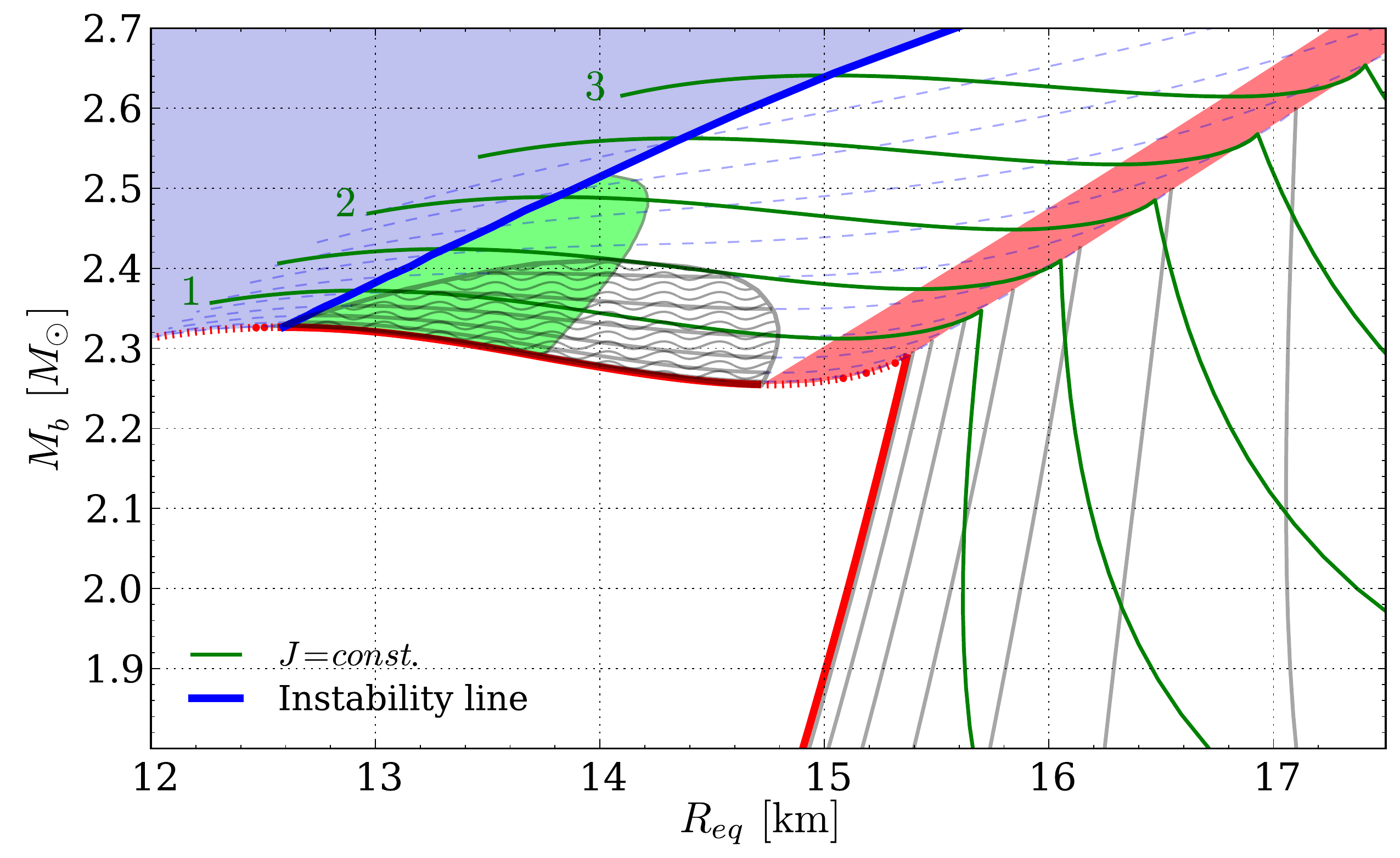}}
\caption{(Color online) Stable and unstable regions on the baryon
mass--equatorial radius $M_{\rm b} - R_{\rm eq}$ plane. The lowest (red) line denotes the
nonrotating configurations. The wavy pattern region denotes a region in which
the back-bending phenomenon does not occur, whereas dashed lines are constant
frequency tracks in a region in which back-bending occurs (lines every 100 Hz,
in the range 100-1000 Hz). The solid green lines denote sequences of configurations
with constant angular momentum $J$ (1,1.5,$\dots$,3, in units of
$GM^2_\odot/c$). The thick solid blue line on the left side marks the boundary
of the region in which rotating axisymmetric stars are unstable with respect to
axisymmetric perturbations. The red strip on the right is the instability
induced by the EOS, dividing the $M_{\rm b} - R_{\rm eq}$ plane into two disjoint
families of solutions.  Finally, the green region is the allowed twin branch
part of the $M_{\rm b} - R_{\rm eq}$ plane to where NSs collapse after entering the
instability strip (assuming $M_{\rm b}=const.$ and $J=const.$).}
\label{fig:mbr_regions}
\end{figure}
%-----------------------------------------------------------------------

Stationary uniformly rotating configurations are labeled (determined) by two
parameters. In Fig.\;\ref{fig:mbr_regions} we show  various continuous
two-parameter curves $[M_{\rm b}(\lambda_{\rm c},\beta), R_{\rm
eq}(\lambda_{\rm c},\beta)]$  in the $M_{\rm b}-R_{\rm eq}$ plane. Here, the
parameter  $\lambda_{\rm c}$ can for instance be the central pressure $P_{\rm c}$, or
central baryon chemical potential $\mu_{\rm b(c)}$ (both behaving continuously
and monotonously along the curve). The quantity $\beta$ characterizes the uniform
rotation of the star; we chose it to be equal to the frequency of rotation $f$  or the total
stellar angular momentum $J$. In Fig.~\ref{fig:mbr_regions} we study the
regions of back-bending and stability on the baryon mass $M_{\rm b}$-equatorial
radius $R_{\rm eq}$ plane. The region in which the back-bending phenomenon is
present is marked with the dashed $f=const.$ lines. An isolated NS ($M_{\rm
b}=const.$) that decreases its angular momentum (by electromagnetic
dipole radiation, e.g.) crosses the lines of $f=const.$ while moving from the right to the 
left side of Fig.~\ref{fig:mbr_regions}. According to Eq.~(\ref{eq:back-bend}),
in the region of dashed lines, the $M_{\rm b}=const.$ line crosses the
$f=const.$ curves such that it results in a {\it \textup{spin-up}} while the angular
momentum is monotonically decreasing, that is, the back-bending. The wavy pattern
region denotes the opposite, usually observed behavior, that is, {\it \textup{spin-down}} with
angular momentum loss.  

Figure~\ref{fig:mbr_regions} also shows two instability regions. The first
is the familiar instability with respect to axisymmetric perturbations related
to the existence of the maximum mass; the blue line corresponding to maxima of
$J$ and denotes its boundary.  The second instability is induced by the strong
phase transition (red strip between the local minimum and maximum of $J$). The
latter divides the space of stable solutions into two disjoint families at any
rotation rate \citep{Zdunik2006}.  Assuming that at some moment in its evolution 
an isolated NS enters the instability strip, it becomes then unstable
and collapses to another, more compact configuration along the lines of $M_{\rm
b}=const.$ and $J=const.$ \citep{Dimmelmeier2009}. 
 
\begin{figure}%[ht]
\resizebox{\columnwidth}{!}
{\includegraphics{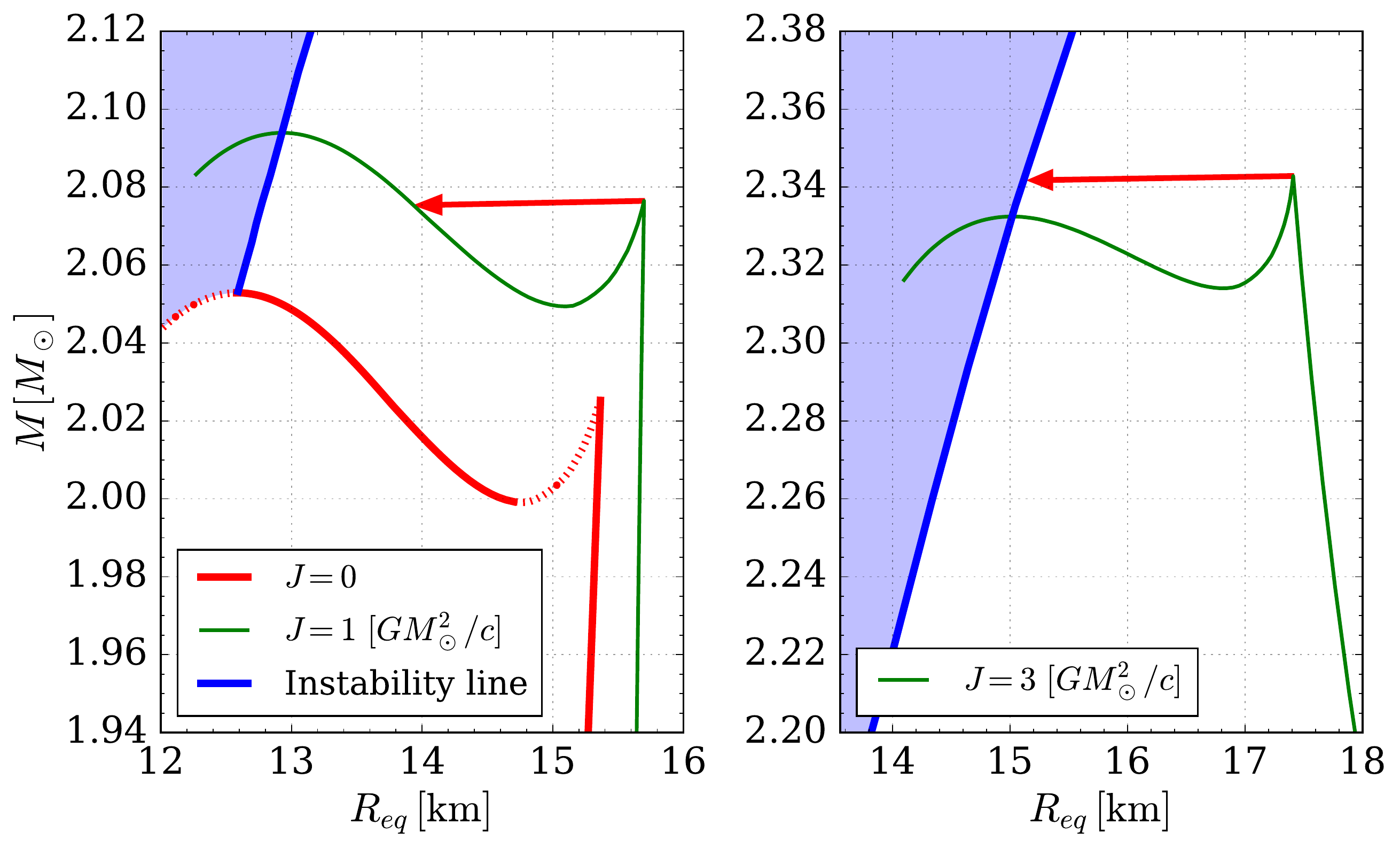}}
\caption{(Color online) Catastrophic mini-collapse for two configurations with
different values of angular momentum $J$ at the edge of the instability region, 
on the gravitational mass--equatorial radius $M - R_{\rm eq}$ plane. 
For an isolated star, the left panel shows the trajectory along a line of constant baryon 
mass $M_{\rm b}$ (note the arrow is inclined, since the ordinate 
is $M$, not $M_{\rm b}$) to another stable NS configuration with $J$ equal to the initial
one.  Right panel: if the angular momentum is larger than some critical value
that depends on the EOS, the star collapses directly to a BH, since the mass of the twin
branch counterpart is too low.}
\label{fig:mbr_jumps}
\end{figure}

Configurations that survive such a mini-collapse are located in the green
region. We note that in this idealized picture no mass and
angular momentum loss is assumed.  Realistically, some mass and angular
momentum loss may occur, and so the green region will decrease toward lower
spin rates and toward the region that is stable against the back-bending, which is marked with
the wavy pattern in Fig.~\ref{fig:mbr_regions}. Careful analysis of 
Fig.\;\ref{fig:mbr_regions} reveals the existence of a critical angular 
momentum $J_{\rm crit}$, such that for $J>J_{\rm crit}$ exceeding the  maximum 
mass on the B branch (resulting from accretion) implies in a direct collapse 
into a rotating BH. This situation is depicted in detail in 
Fig.~\ref{fig:mbr_jumps}. 

%-----------------------------------------------------------------------
%-----------------------------------------------------------------------

\begin{figure}%[ht]
\resizebox{\columnwidth}{!}
{\includegraphics{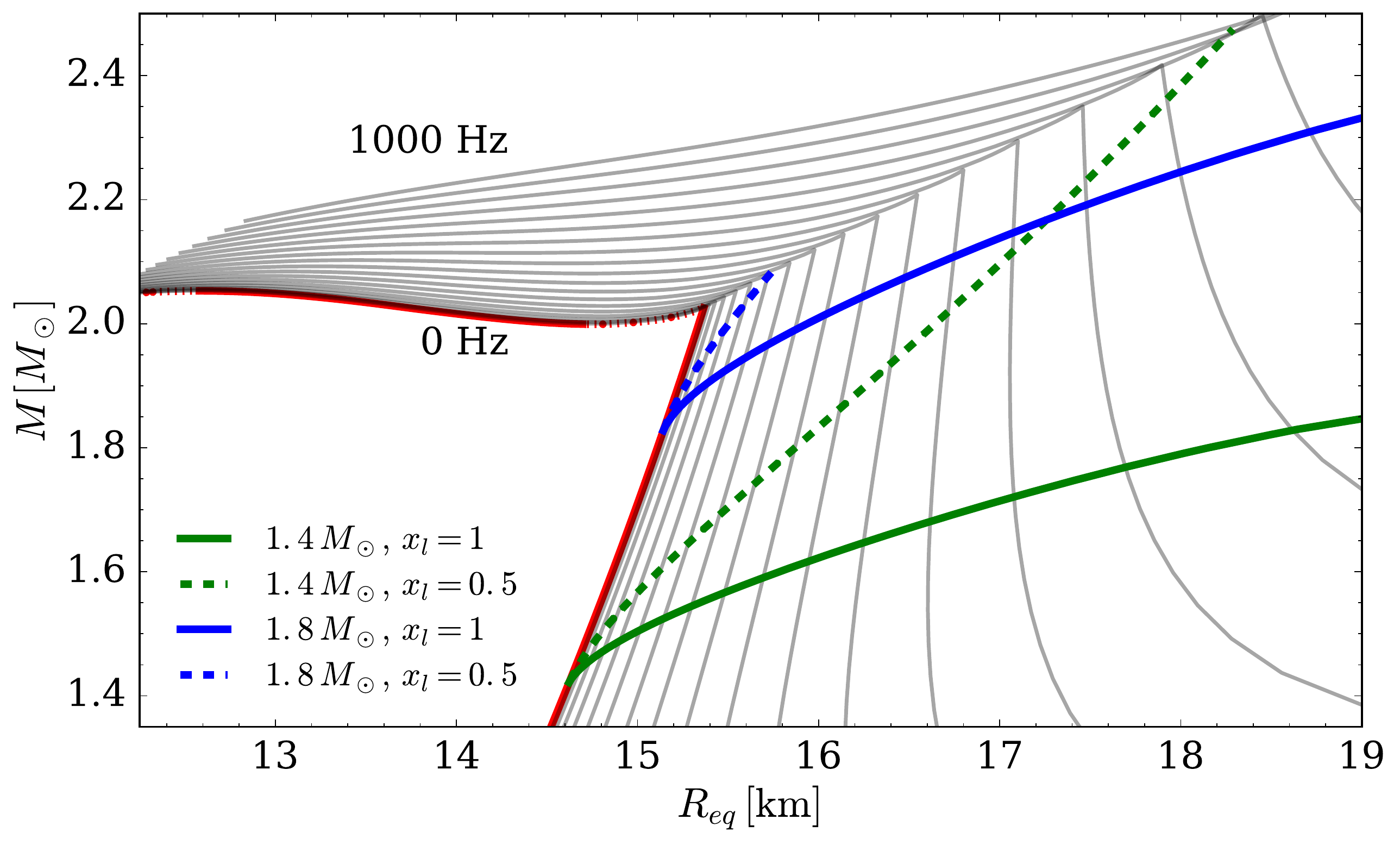}}
\caption{(Color online) Evolutionary tracks on the gravitational
mass--equatorial radius $M - R_{\rm eq}$ plane, assuming a thin-disk accretion
model, superimposed on the mass-radius curves for constant rotation rates in
the range between 0 and 1000 Hz. Two angular momentum transport efficiencies
are considered. Solid lines correspond to perfect angular momentum transfer
from the accretion disk to the star, dashed lines correspond to half of the
accretion-disk angular momentum transferred.  The initial masses for these
tracks are $1.4\;M_\odot$ and $1.8\;M_\odot$. For more details see the text.}
\label{fig:mr}
\end{figure}

The existence of $J_{\rm crit}$ and its potential observational signatures are discussed in 
more detail in Sect.~\ref{sect:discussion}. Here we restrict ourselves to a few 
comments on how the existence of $J_{\rm crit}$ is due to generic features of the 
rotating-twin $M(R_{\rm eq})$ curves. Consider first the static ($J=0$) B and Q branches. 
They have very different $M(R)$ dependence. The maximum mass  at 
$M=M^{\rm (B)stat}_{\rm max}$ (with $P_{\rm c}=P_{_{\rm BQ}}$) is due is to the strong phase 
transition, and is  somewhat lower than the flat maximum on the Q branch, 
$M^{\rm (Q)stat}_{\rm max}$. We proceed to the case of an increasing $J$, when  
an NS on the B branch acquires mass and angular momentum from an accretion disk. 
At some moment it reaches a maximum on the B branch; it has then $J=J_1$, 
$M^{(\rm B)}_{\rm b}=M_{\rm b1}$.  Under further accretion it collapses into a 
Q configuration of the same $M_{\rm b1}$ and $J_1$, provided  
$M_{\rm b1}<M_{\rm b,max}^{(\rm Q)}(J_1)$. The equality is reached exactly at $J_{\rm crit}$, 
and for higher $J$ the B star collapses directly into a BH. A critical $J$ is equivalent 
to maximum $f,$ however, which can be reached by the baryon stars. This situation is depicted in detail 
in Fig.~\ref{fig:mbr_jumps}  and is discussed further in Sect.~\ref{sect:discussion}. 
A separate question is related to the way in which the NS reaches the instability line. 
As shown in \citet{Zdunik2005} and \citet{Bejger2011}, the efficiency of transfer of the angular 
momentum in the process of disk accretion governs the $M(R)$ evolution of a spinning-up 
NS. To illustrate how an NS can reach the unstable region, we present in Fig. 3 the mass--equatorial radius dependence
for evolutionary tracks of accreting NS.  A realistic evolutionary path depends
on many physical details, such as the configuration and strength of the magnetic field,
the accretion rate, and the way the magnetic field interacts with the disk.
We assume thin-disk accretion in the presence of the magnetic field by employing a
model used previously in \citet{Bejger2011a} and \citet{Fortin2016a}. Two
different efficiencies of angular momentum transfer, $x_l=0.5$ and $x_l=1$, are
considered for initially nonrotating configurations of mass $1.4\,M_\odot$ and
$1.8\,M_\odot$. The initial magnetic field is $B=10^8$ G. If the accretion
stops before the instability is reached, the star evolves by moving
horizontally from right to left (secular spin-down). Figure\;\ref{fig:mr} is
intended to show that it is possible to reach the instability region at low or
high angular momentum corresponding to the two cases considered in
Fig.\;\ref{fig:mbr_jumps}.

\begin{figure}[ht]
\resizebox{\columnwidth}{!}
{\includegraphics{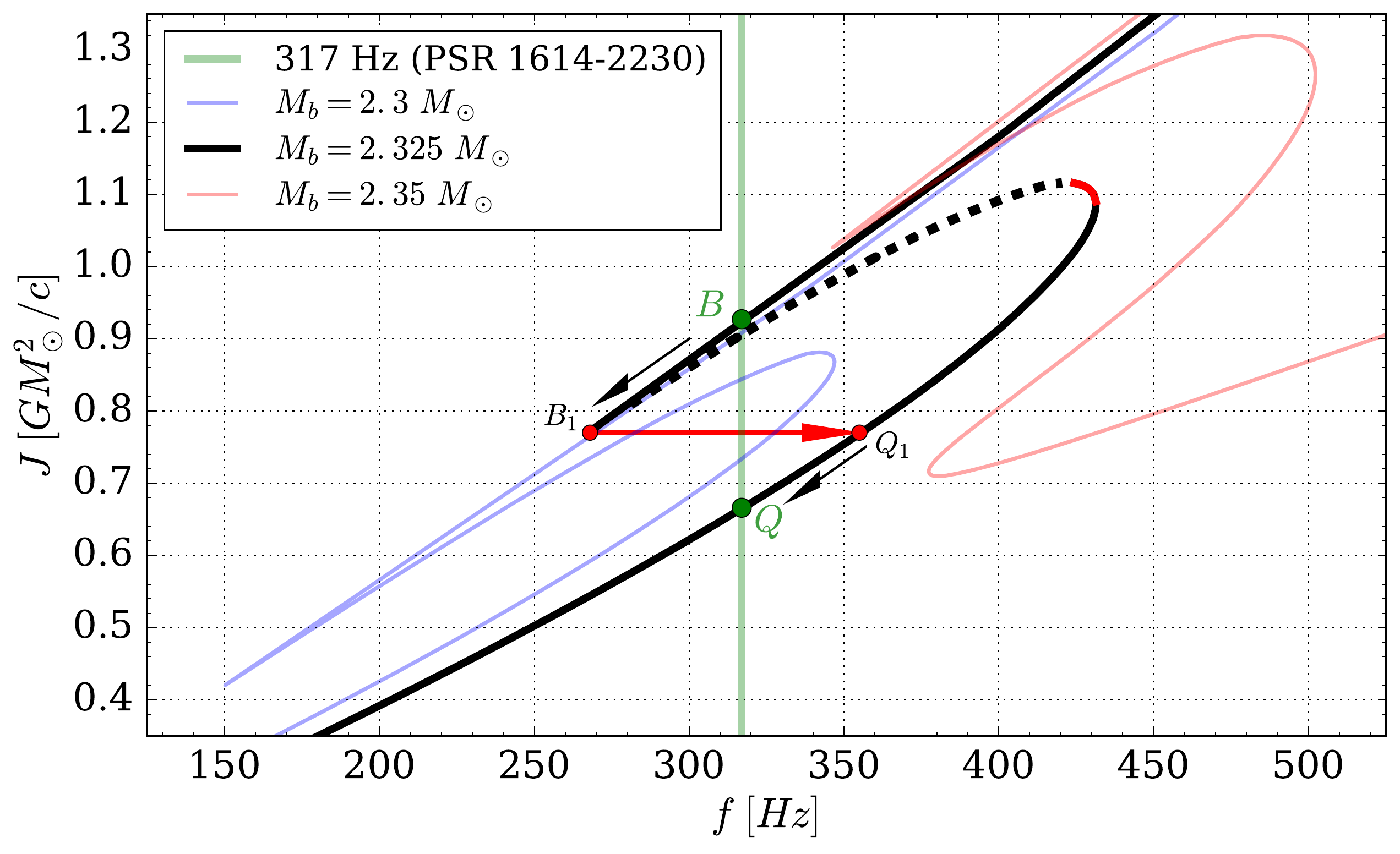}}
\caption{(Color online) Evolution of a spinning-down NS with a 
constant baryon mass $M_{\rm b}$ on the $J - f$ plane for selected values 
of $M_{\rm b}$: $2.3\ M_\odot$ (blue line), $2.325\ M_\odot$ (black line), 
and $2.35\ M_\odot$ (red line). The thick red segment on the $2.325\ M_\odot$ 
sequence denotes the back-bending region, which is never reached 
during the angular momentum loss evolution. Superimposed, the spin frequency 
of the massive NS PSR 1614-2230. For a detailed description see the text.}
\label{fig:jf}
\end{figure}

The effect of the EOS-induced instability strip may also be studied on the
angular momentum--spin frequency $J - f$ plane, assuming a fixed baryon mass
sequence, see Fig.~\ref{fig:jf}. The evolution of an isolated NS that
decreases its angular momentum $J$ corresponds to a downward movement in this
figure. This also means that the central density (and pressure) of such a star
increases. A spinning-down NS on the B branch decreases its $J$ until it
becomes unstable (point B$_1$), which forces it to collapse (dynamically
migrate) to another stable branch (point Q$_1$) with the same $M_{\rm b}$ and
$J$.  We note here that in principle by adopting a certain EOS we may place
constraints on the parameters of massive pulsars with known spin frequency. In the
example from Fig.~\ref{fig:jf}, an NS with $M_{\rm b}=2.325\; M_\odot$ and the
spin frequency of PSR 1614-2230 (317 Hz) is located on either the upper
baryonic branch denoted by B, or on the lower branch (hybrid stars - Q), the two
configurations differing greatly in the total angular momentum because of very
different moments of inertia. A configuration with a slightly lower $M_{\rm
b}=2.3\; M_\odot$ (blue line) may, for this particular EOS, exist at 317 Hz
only in the B phase, since the instability transfers it to frequency
${\simeq}200$ Hz (significantly lower than 317 Hz). A configuration with a slightly
higher $M_{\rm b}=2.35\; M_\odot$ (red line) is excluded as a model of PSR
1614-2230; during its evolution, it never spins down to reach 317 Hz: after the
dynamical migration and a period of spin-down, it will collapse to a BH. 

\begin{figure}[ht]
\resizebox{\columnwidth}{!}
{\includegraphics{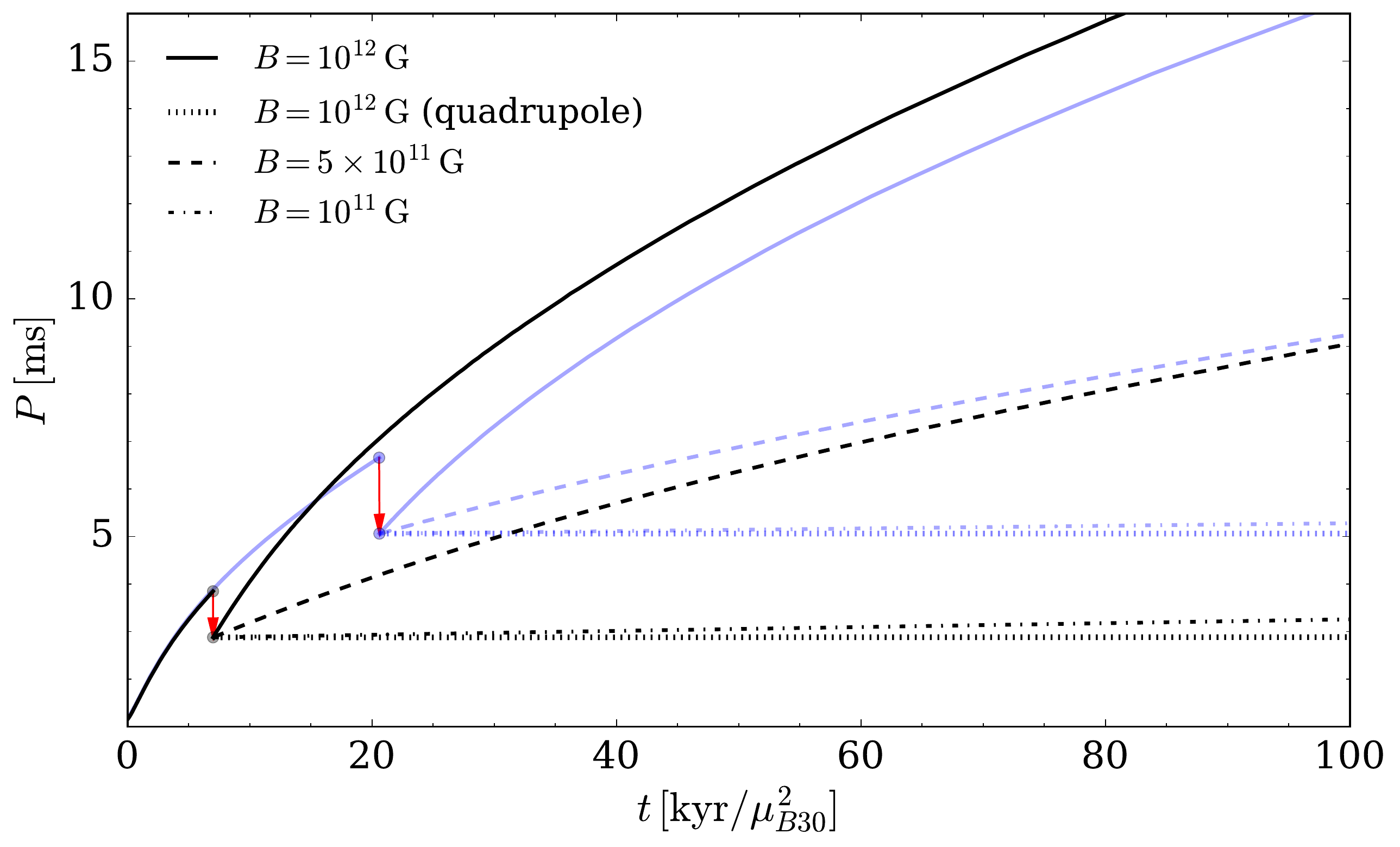}}
\caption{Evolution of the spin period of a spinning-down NS with constant
baryon mass $M_{\rm b}=2.325\ M_\odot$ (black lines) and $M_{\rm b}=2.3\
M_\odot$ (blue lines), in relation to Fig.~\ref{fig:jf}. The time unit is normalized by
the value of the magnetic dipole moment $\mu_B=BR^3=10^{30}$~[cgs], which
corresponds to a magnetic field of $B=10^{12}$ G and a stellar radius of $R=10$
km. The lines after the mini-collapse related to a sudden spin-up and marked
with red arrows are denoted according to the assumption that the $B$ field
stayed unchanged (solid lines), decreased twice (to $B=5\times 10^{11}$ G,
dashed lines), or decreased by one order of magnitude (to $B=10^{11}$ G,
dash-dotted lines). Dotted lines mark the evolution of a system with a purely 
quadrupolar magnetic field of $B=10^{12}$ G, emulating a drastic change 
of the field structure after the mini-collapse.}
\label{fig:tt}
\end{figure}

The time evolution of the rotation period of an isolated NS loosing its energy
and angular momentum due to the dipole radiation is presented in
Fig.~\ref{fig:tt}. The assumption of dipole radiation from the pulsar leads to the
formula
\begin{equation}
 \frac{{\rm d}E}{{\rm d}t}=-\frac{\mu^2_B\Omega^4\sin^{2}\alpha}{6c^3}, 
\end{equation}
where $E$ is the energy of rotating pulsar and $\mu_B=B\,R^3$ the dipole moment of
a star. In the framework of general relativity, we should use the total
mass-energy of the star $Mc^2$ in place of $E$. For the evolution of an isolated NS
with a fixed total number of baryons, the relation ${\rm d}M=\Omega/c^2\,{\rm d}J$
holds, resulting in the equation for the time evolution, 
\begin{equation}
\frac{{\rm d}\tilde{J}}{{\rm d}\tilde{t}}=-3.34\times 10^{-9}{B_{12}^2\,R_{6}^6\,f_{_{\rm Hz}}^3~,} 
\end{equation}
where $\tilde{J}$ is the stellar angular momentum in the units of $GM^2_\odot/c$ and $\tilde{t}$ is time in kyr 
(the same relation in a Newtonian case could be obtained for
$E=\frac{1}{2}I\Omega^2$ and $J=I\Omega$, with $I$ being the NS moment of inertia).  
The timescales of slowing down of a solitary pulsar from milliseconds (close to 
maximum mass-shedding frequency) down to ${\sim}15$~ms before and after the mini-collapse
are clearly comparable - we assumed that the magnetic moment $\mu_B$ does either not change 
during the evolution, or that the magnetic field decreases. 

%-----------------------------------------------------------------------
%-----------------------------------------------------------------------
%-----------------------------------------------------------------------
%-----------------------------------------------------------------------

\begin{figure}[ht]
\resizebox{\columnwidth}{!}
{\includegraphics{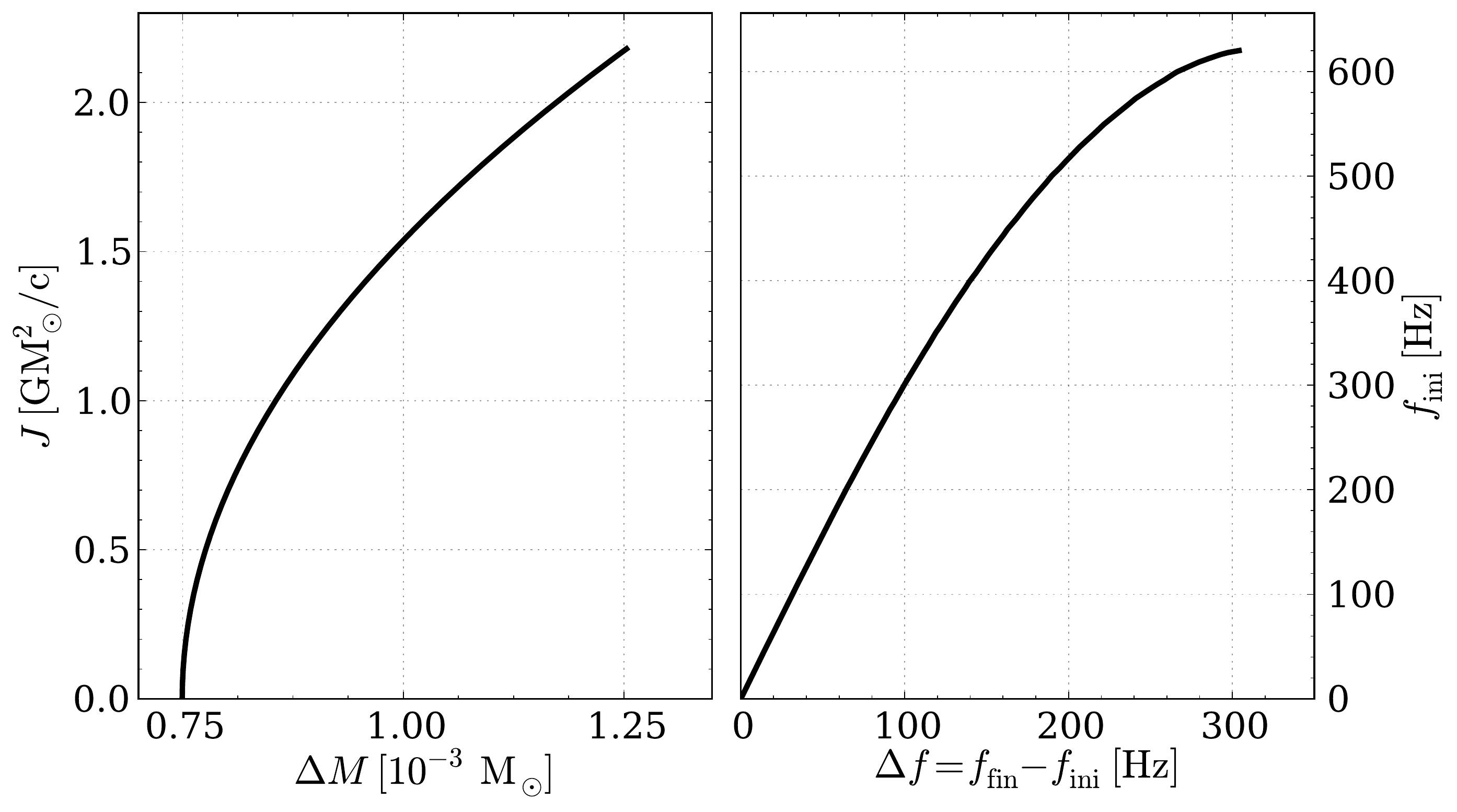}}
\caption{Left panel: energy release (difference in 
the gravitational mass) against the angular momentum of the configuration 
entering the strong phase transition instability. Right panel: spin-up 
(difference between the final and initial spin frequency) against 
the spin frequency of the initial configuration. For details see 
the text.}
\label{fig:j_dm_f_df}
\end{figure}

A mini-collapse is a dynamical process, provided the B$\longrightarrow$Q
conversion has detonation character \citep{Haensel2016}. It involves
considerable spin-up and a substantial reorganization of the interior of the
star \citep{Dimmelmeier2009}. It is also exoenergetic: a substantial amount of
energy, quantified as the difference between the initial and the final
gravitational mass, $\Delta M = M_{\rm ini} - M_{\rm fin}$, is released in the
process. The left panel of Fig.~\ref{fig:j_dm_f_df} shows the relation between the
angular momentum $J$ of the star and the energy difference $\Delta M$.  The
$\Delta M(J)$ relation is approximately quadratic in $J$: $\Delta M(J) = aJ^2 + \Delta
M(0)$. For the particular EOS used in this study, $a=0.106$ and $\Delta
M(0)=0.749$, for $\Delta M$ in $10^{-3}\ M_\odot$ and $J$ in $GM^2_\odot/c$
units. The line ends at a critical $\tilde{J}\approx 2.20,$ where the
dynamical collapse forces an NS to the instability region, where it collapses to
a BH (see also the right panel of Fig.~\ref{fig:mbr_jumps}).  The right panel
of Fig.~\ref{fig:j_dm_f_df} shows the amount of spin-up (spin frequency
difference) $\Delta f = f_{\rm fin} - f_{\rm ini}$ that is acquired in the
mini-collapse as a function of the initial spin frequency $f_{\rm ini}$. The
situation presented here corresponds to a specific case studied in detail in
\citet{Zdunik2008}, where the overpressure of the new metastable phase is set
to zero, $\delta\bar{P}=0$. 

From the same plot one may estimate the `Newtonian' change of kinetic energy and 
the luminosity of the process. Assuming that the total angular momentum does 
not change during the process, we obtain 
\begin{equation} 
\Delta E^{\rm rot} = E^{\rm rot}_{\rm fin} - E^{\rm rot}_{\rm ini} = \frac{1}{2}J\left(\Omega_{\rm fin} - \Omega_{\rm ini}\right).
\label{eq:delta_erot}
\end{equation} 
For the value of $J = 2\,{GM_\odot^2/c}$ the change of the spin frequency is approximately 
$\Delta f = 240$ Hz ($\Delta \Omega = 1510$ rad/s). 
For these figures we obtain $\Delta E^{\rm rot}\simeq 2\times 10^{52}$ erg. 
This value overestimates the difference in total gravitational mass $\Delta M$ 
by one order of magnitude ($10^{-3}\, M_\odot c^2\simeq 2\times 10^{51}$ erg).  
The process occurs on a dynamical timescale of a millisecond. 

%-----------------------------------------------------------------------
%-----------------------------------------------------------------------
%-----------------------------------------------------------------------
\begin{figure}[ht]
\resizebox{\columnwidth}{!}
{\includegraphics{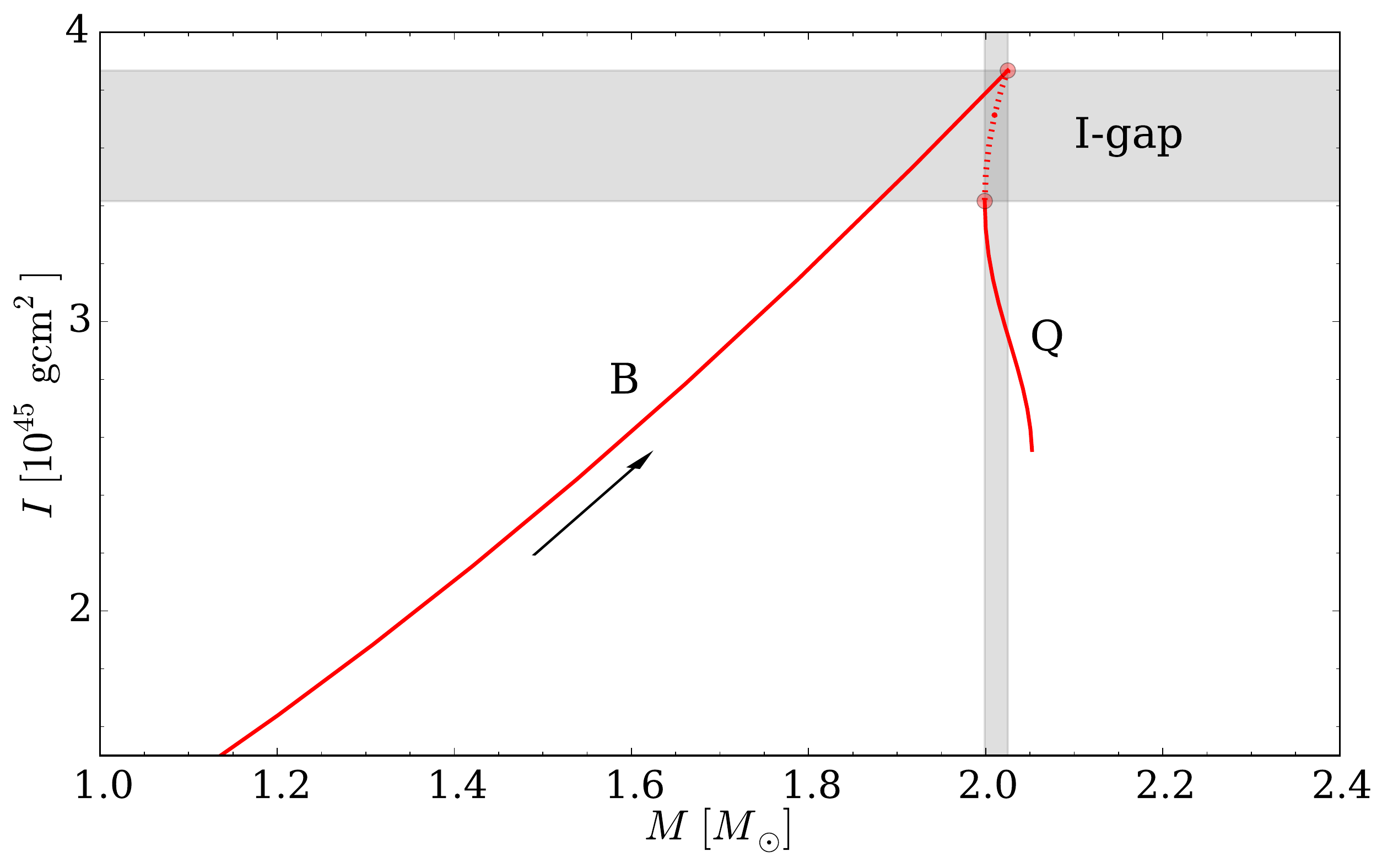}}
\caption{Stellar moment inertia $I$ versus gravitational mass $M$ for nonrotating NS models. Solid
lines: stable equilibrium configurations. The arrow indicates increasing central
density. The upper filled circle corresponds to the termination of the stable B branch, and
the lower circle correspond to the first stable equilibrium configuration of the
Q branch. The Q-branch line ends at the last stable equilibrium configuration. 
The dotted line corresponds to unstable configurations. 
The $I$ - gap between the stable B and Q twin branches is shown.
For details see the text.}
\label{fig:I-M-R}
\end{figure}
%-----------------------------------------------------------------------

%///////////////////////////////////////////////////////////////////////
\section{Discussion}
\label{sect:discussion}
%///////////////////////////////////////////////////////////////////////
The goal of this article is to {\it \textup{bona fide}} consider a specific class of EOS
featuring a substantial phase transition motivated by the theory of dense
matter physics. The EOS-induced instability region divides stable NS
configurations into two disjoint families (twin families). Its existence has
interesting astrophysical consequences for rotating NSs. We note that it 
facilitates a natural (i.e., not fine-tuned) way for various astrophysical phenomena that we list below. 
\parindent 0pt
\vskip 4mm
{\it Spin frequency cutoff}. Even though theoretical models of NSs
allow for spin rotation rates much above 1 kHz and although with current observational 
techniques such rapidly rotating pulsars could be detected (see, e.g., \citealt{P10,DE11}), so far,  
the most rapidly rotating NS
observed is PSR J1748-2446ad (716 Hz, \citealt{Hessels2006}).
It cannot be excluded {\it \textup{a priori}} that some rapidly rotating and
massive NSs were created close to their currently observed state, that is, in a
specific type of core-collapse supernov\ae. If this were the
case, then they might
appear practically everywhere on the right side of the thick blue line of
Fig.~\ref{fig:mbr_regions}, with the exception of the instability strip (red
area), where no stationary axisymmetric solutions are possible. 
However, as the observations, evolutionary arguments, and numerical
simulations tend to suggest, NS that become radio pulsars are not born with
${\sim}$1 - 3 ms periods, but with much longer periods of ${\sim}20-150$ ms, 
see \citet{FGK06}, \citet{Kramer2003}, Table 7.6 in 
\citet{LyneG1998}, and references therein (a
specific class of NS with millisecond periods at birth in massive core-collapse
supernovae are thought to be progenitors of {\it \textup{magnetars}}, which are observed as
soft-gamma ray repeaters or anomalous X-ray pulsars, see e.g.,
\citealt{KargaltsevPavlov2008}). Then, after slowing down 
to  a period of a few seconds and entering the pulsar graveyard, they gain their angular
momentum, as well as mass, during long-term accretion processes in low-mass
binary systems (in the so-called recycling  of dead pulsars, see
\citealt{Alpar1982,Radhakrishnan1982,Wijnands1998}), and it is possible
 that some of them enter the strong phase transition instability strip
sometime in their evolution. Sufficiently massive and sufficiently rapidly
rotating NS will then migrate dynamically along the $M_{\rm b}=const.$ track in
the direction of the twin branch (see, e.g., \citealt{Dimmelmeier2009}).
Moreover, for some critical angular momentum (critical spin frequency) the
value of $M_{\rm b}$ on the right side of the instability strip (the $M_{\rm
b}(R_{\rm eq})$ peaks in Fig.~\ref{fig:mbr_regions}) is higher than the corresponding
maximum of $M_{\rm b}$ on the twin branch 
- in that case, the star collapses to a BH. 
\vskip 2mm
{\it No observations of back-bending in radio-pulsar timing}. 
One of the observational 
predictions related to substantial dense-matter phase transitions
is the detection of the back-bending phenomenon, which occurs at spin frequencies
of known pulsars. As we showed in Sect.~\ref{sect:results},
Fig.~\ref{fig:mbr_regions}, NSs exhibiting an instability that
is caused by a strong phase
transition avoid the vast majority of the back-bending region for spin
frequencies lower than some critical value. The most rapidly
rotating currently
known pulsar, PSR J1748-2446ad, has a spin period of 716 Hz. From
Fig.~\ref{fig:mbr_regions} we note that the $f=700$ Hz line is the first
dashed line above the no back-bending (wavy pattern) region. When we assume that
the EOS used for illustration is the true EOS of dense matter, this means that PSR
J1748-2446ad, which does not show the features of back-bending in the timing,
still resides on the hadronic branch (does not contain the quark core).
Additionally, the most massive stars that are in the back-bending region may
not be effective pulsars - they may be electromagnetically exhausted, with
their magnetic field dissipated in the violent process of mini-collapse, 
and therefore not easily detectable. 

\vskip 2mm
{\it Radius gap}. The existence of an instability strip creates in a mass- and spin-frequency-dependent radius region of avoidance between the allowed green
region and stable baryonic branch on the right of Fig.~\ref{fig:mbr_regions},
which is broadened with increasing mass (see also Fig. 3 of
\citealt{Benic2014}). For nonrotating $2\;{\rm M}_\odot$ NS, the predicted
$R$-gap is ${\sim}1$ km. Small-radius Q-branch twins have $R^{\rm (Q)} = 12.5\;
{\rm km}- 14.5\;{\rm km}$ within a very narrow mass range ${\sim}0.1\;{\rm
M}_\odot$. The measurement of a radius $R>15\;$km for a ${\sim}2\;{\rm M}_\odot$ star 
indicates a B-branch configuration. If for another 
${\sim}2\;{\rm M}_\odot$ NS 
the radius is determined to be within the range  $12.5\; {\rm km}- 14.5\;{\rm km}$, then we obtain strong  evidence in favor of distinct B and Q twin branches as a result of a strong
B$\longrightarrow$Q phase transition. 

\vskip 2mm
{\it Moment of inertia gap}. The twins on the B  and Q branches have different
internal structure. The B-star twin is more compact, and its mass is
concentrated in the dense quark core. Consequently, at the same $M$ of twins, one
has  $I^{\rm (Q)}$ significantly smaller than $I^{\rm (B)}$ ,
as shown on Fig.~\ref{fig:I-M-R}.
  Moreover, the strong first-order phase transition
B$\longrightarrow$Q results in an $I$ gap  between the B and Q twins
(Fig.~\ref{fig:I-M-R}). The moment of inertia of NS can be measured through 
the spin-orbit effect contribution to the timing parameters for  a binary of two
radio pulsars \citep{DamourSchaefer1988}. The first binary of this type, 
PSR J0737-3039A,B was discovered  more than a decade ago \citep{Lyne2004}. 
Pulsar B has become invisible to terrestrial observers in March 2008 because
its beam wandered out of our line of sight as a consequence of the geodetic precession
effect \citep{Perera10}. It may reappear as late as in 2034 (or later),
depending on the model of the pulsar magnetosphere (see,
e.g., \citealt{LL14}). Since the mass of pulsar A has been accurately
determined, a measurement of its moment of inertia through the spin-orbit
momentum coupling would allow us to constrain the
radius and hence the EOS \citep{LattimerSchutz2005}. Given the present timing
accuracy of the system's post-Keplerian parameters, that is, the periastron
advance, the decrease in the orbital period and the Shapiro shape parameter,
from which the spin-orbit coupling contribution is derived, reasonable accuracy may be
achieved around the time of pulsar B reappearance in ${\simeq}20$ years
\citep{Kramer09}. However, in the forthcoming era of large radio telescopes
(e.g., FAST, SKA) the number of known pulsars will increase by orders of
magnitude, including many thousands of millisecond pulsars, out of which we may
hopefully expect tens of binary systems with two pulsars suitable for
simultaneously measuring  $I$ and $M$ of an NS. A sufficiently dense set of
pairs  $\lbrace I_i,M_i \rbrace$ resulting from these  future measurements
could then be used to confirm or reject the generic $I(M)$ shape  in
Fig.\;\ref{fig:I-M-R}. 

\vskip 2mm
{\it EOS-induced dynamical collapse as an energy reservoir}. 
We assume that the B$\longrightarrow$Q phase transition, mediated by strong
interaction, occurs in the detonation regime. A dynamical process involving
NS, triggered by the loss of stability, is associated with a substantial energy
release (${\sim}10^{52}~{\rm erg}$), heating of dense matter, kinetic energy
flow, and some emission of radiation, on both a short time-scale (NS quake,
mini-collapse) and long time-scale (surface glowing).  The process is probably
also associated with a substantial rearrangement  of the NS magnetic field. For
NS with total angular momenta larger than some critical value, it leads to a
direct collapse to a BH. This event is related to the expulsion of the
magnetic field and thus the dynamical migration to the high-mass twin branch
may be considered a natural extension of the 
\citet{FalckeR2014} cataclysmic scenario. Alternatively, for NSs below the angular-momentum
threshold, the mini-collapse dynamics may influence the magnetic field to such
an extent that it becomes a transient source of observable magnetospheric 
emission after the final configuration ends up on the twin branch. 
\vskip 2mm
\parindent 21pt

To conclude, the assumption of the strong phase transition in the NS EOS leads
to a number of falsifiable (at least in principle) astrophysical predictions.
As we described in Sect.~\ref{sect:results} and in Fig.~\ref{fig:jf}, a
transition to a new exotic phase (deconfined quarks in this example) constrains
the range of available $M_{\rm b}$ and $J$. This reasoning can be extended to
other EOS functionals, like the moment of inertia $I$; moreover, these
constraints may be combined with the observations that are sensitive to the composition
of the core, for example, NS  cooling studies.  The evolutionary scenario in which some
of NS collapse to BHs produces a specific NS-BH mass function without a mass
gap, which should be possible to test with current and future searches for
low-mass BHs with microlensing surveys \citep{Wyrzykowski2016},
for instance. 
Unless the NS magnetic field is amplified and/or
reoriented during the mini-collapse event, it is likely that it is dissipated
and disordered during the process - in the latter case, we expect
a population
of massive ineffective pulsars with a low magnetic field. Moreover, a dynamical
mini-collapse creates a characteristic signature of the GW 
emission, strongly dependent on the EOS and mass of the NS; short
transient GW radiation of this type should be detectable by the advanced
era interferometric detectors \citep{Dimmelmeier2009}. If the collapse is not entirely
axisymmetric, the final configuration may retain an asymmetry, thus creating a
rotating NS that spins down while continuously emitting the
almost-monochromatic GWs. Such objects are among the prime astrophysical
targets of the Advanced LIGO and Advanced Virgo interferometric detectors (see
the demonstration of search pipelines used on the initial LIGO and initial Virgo 
detector data, e.g., \citealt{Aasi2014CQGra,Aasi2014PhRvD,Aasi2015ApJ}). 

There are several open questions related to the various aspects of the
high-mass twin scenario, such as what the conditions are for the NS to reach the
instability region via the disk accretion spin-up, what the influence of the
possible meta-stability of the quark phase core is on the NS population on the
twin branch, and how the electromagnetic and gravitational-wave emission depends 
on the parameters of the NS. These questions will be addressed in subsequent studies.

%///////////////////////////////////////////////////////////////////////
\begin{acknowledgements}
This work has been supported in part by the Polish National Science Centre
(NCN) under grant No. UMO-2014/13/B/ST9/02621. D.B. has been supported in part 
by the MEPhI Academic Excellence Project under contract No. 02.a03.21.0005. 
The authors acknowledge support from the COST Action MP1304 ''NewCompStar''
for their networking activities.  
\end{acknowledgements}

%-----------------------------------------------------------------------

\end{document}